\documentclass[epjCONF]{svjour}
\usepackage{graphicx}
\usepackage[varg]{txfonts} 
\usepackage[latin1]{inputenc}
\session-title{%
19$^{\textnormal{\footnotesize th}}$ International %
IUPAP Conference on Few-Body Problems in Physics%
}
\begin{document}
\title{%
Kaon-Nucleon potential from lattice QCD
}%
\author{%
Y. Ikeda\inst{1}\fnmsep\inst{2} \fnmsep\thanks{\email{yikeda@nt.phys.s.u-tokyo.ac.jp}} 
\and %
S. Aoki\inst{3}
\and %
T. Doi\inst{3} 
\and %
T. Hatsuda\inst{1} 
\and %
T. Inoue\inst{3} 
\and %
N. Ishii\inst{1}
\and %
K. Murano\inst{3} 
\and %
H. Nemura\inst{4} 
\and %
K. Sasaki\inst{3}
\includegraphics[width=0.5\columnwidth]{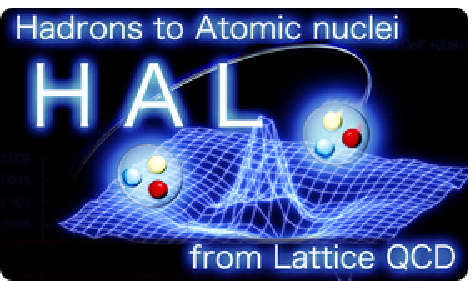}
}
\institute{%
Depertment of Physics, The University of Tokyo, 
Tokyo 113-0033, Japan.
\and %
Nishina Center for Accelerator-Based Science, Institute for Physical
and Cemical Research (RIKEN), \\ Wako, Saitama 351-0198, Japan
\and %
Depertment of Pure and Applied Sciences, The University of Tsukuba, 
Tsukuba, Ibaraki 305-8577, Japan
\and %
Depertment of Physics, Tohoku University, 
Sendai, Miyagi 980-8578, Japan
}
\abstract{
We study 
the $KN$ interactions in the 
$I(J^{\pi})=0(1/2^-)$ and  $1(1/2^-)$ channels 
and associated exotic state $\Theta^+$ 
from 2+1 flavor full lattice QCD simulation for relatively
heavy quark mass corresponding to $m_{\pi}=871$ MeV.
The s-wave $KN$ potentials are obtained from the Bethe-Salpeter
wave function by using the method 
recently developed by HAL QCD (Hadrons to Atomic nuclei from Lattice QCD)
Collaboration.
Potentials in both channels reveal short range repulsions:
Strength of the repulsion is stronger in the  $I=1$ potential, which is 
consistent with the prediction of the Tomozawa-Weinberg term.
The $I=0$ potential is found to have attractive well at mid range. 
From these potentials, the $KN$ scattering phase shifts are calculated
and compared with the experimental data. 
} 
\maketitle
%
%
%
\section{Introduction}
\label{SchmidtPL_intro}
First evidence of the $\Theta^+(1540)$
(an exotic state with baryon number $B=1$ and strangeness $S=+1$ which
corresponds to the quark content $uudd\bar{s}$) 
has been reported by the LEPS Collaboration at SPring-8~\cite{LEPS:03}.
Although numbers of experimental studies have been performed since then,
the existence of the $\Theta^+(1540)$ is still controversial:
CLAS Collaboration observed no signal 
from their high statistic data~\cite{CLAS:06},
and  other experiments at high energies
with high statistics and good mass resolutions
did not find positive evidence as reviewed in \cite{Danilov:2007bp}.
On the other hand,
LEPS Collaboration and DIANA Collaboration have
recently reconfirmed the $\Theta^+(1540)$ signal 
with high statistics~\cite{LEPS:08,Barmin:2009cz}.
They indicate that the production mechanism of the $\Theta^+(1540)$ might be highly
reaction dependent if it exits in nature.

Theoretically, QCD studies of the mass and the quantum numbers of the $\Theta^+(1540)$
have been attempted using lattice QCD simulations
and QCD sum rules. 
In the lattice QCD studies with
the quenched approximation,
existence of the low-mass pentaquark
is not conclusive yet (see e.g. the summary given in \cite{Doi:2007sc}).
Recent QCD sum rule studies~\cite{GJKNO:09a} 
with the operator product expansion up to dimension 14
show some candidates of the $\Theta^+(1540)$ in the
$I(J^{\pi})=0(1/2^-)$, $1(1/2^-)$, $0(3/2^+)$ and $1(3/2^+)$ channels.
Determination of the quantum numbers of $\Theta^+(1540)$ was also attempted
from the phase shift analyses of the $KN$ scattering data,
and some candidates in $L_{2I,2J}=D_{03}, F_{05}$ 
were reported ~\cite{KNK:03}.
 
The main purpose of the present study is to investigate the 
low energy $KN$ potentials in the 
$I(J^{\pi})=0(1/2^-)$ and  $1(1/2^-)$ channels 
and to shed  new light on  $\Theta^+$ 
from  the 2+1 flavor full QCD simulations.
A systematic method to extract the hadron-hadron potential 
from the equal-time Bethe-Salpter amplitude measured on the lattice 
has been recently developed and applied to the nucleon-nucleon potential
by HAL QCD  Collaboration~\cite{IAH:06,AHI:09,NIAH:08,NIAH:09,Inoue:09} 
The potential obtained in this method can be utilized to calculate
the scattering observables and to study the resonances and bound states. 
We utilize this method to extract the $KN$ potentials and phase shifts.

This paper is organized as follows.
In section 2, the formalism to extract the $KN$ potential from lattice QCD
is briefly reviewed.
Our numerical setup of the lattice QCD simulation is then shown in section 3,
and the results are shown in section 4.
These results are discussed in section 5, 
and summary is given in section 6.

\section{Formalism}
Following the basic formulation to extract the nucleon-nucleon interaction~\cite{IAH:06,AHI:09},
we briefly show the equations to obtain the $KN$ potentials below.
We start with an effective Schr\"{o}dinger equation for 
the equal-time Bethe-Salpeter (BS) wave function $\phi(\vec{r})$:
\begin{equation}
-\frac{\nabla^2}{2\mu} \phi(\vec{r}) + \int d\vec{r} U(\vec{r},\vec{r'}) \phi(\vec{r'})
= E \phi(\vec{r}),
\end{equation}
where $\mu(=m_K m_N /(m_K + m_N))$ and $E$ denote the reduced mass of the $KN$ system 
and the non-relativistic energy, respectively.
The non-local potential $U(\vec{r},\vec{r'})$ can be expanded
in powers of the relative velocity $\vec{v}=\nabla/\mu$ at low energies,
\begin{eqnarray}
U(\vec{r},\vec{r'}) & = &
V(\vec{r}, \vec{v}) \delta(\vec{r} - \vec{r'})  \nonumber \\
& = & (V_{LO}(\vec{r})+V_{NLO}(\vec{r})+ \cdots) \delta(\vec{r}-\vec{r'}),
\end{eqnarray}
where the $N^n LO$ term is of order $O(\vec{v}^n)$.
In the leading order, we have
\begin{equation}
V(\vec{r}) \simeq V_{LO}(\vec{r}) = \frac{\nabla^2}{2\mu}\phi(\vec{r})+E.
\label{pot}
\end{equation}

In order to obtain the  BS wave function of the $KN$ system on the lattice,
let us consider the four-point correlator:
\begin{eqnarray}
& & {\cal G}_{\alpha}(\vec x, \vec y, t-t_0; J^{\pi}) =
\left\langle 0 \right|
K(\vec x, t) N_{\alpha}(\vec y, t)
\overline{{\cal J}}_{K N}(t_0; J^{\pi})
\left| 0 \right\rangle \nonumber \\
& & \ \ \ \ \ \ \ \  = 
\sum_{n}A_n \left\langle 0 \right|
K(\vec x, t) N_{\alpha}(\vec y, t)
\left| n \right\rangle 
\ e^{-E_n(t-t_0)},
\label{4-point}
\end{eqnarray}
with the matrix elements
\begin{equation}
A_n = \left\langle n \right| 
\overline{{\cal J}}_{K N}(t_0; J^{\pi})
\left| 0 \right\rangle.
\end{equation}
Here $\overline{{\cal J}}_{K N}(t_0; J^{\pi})$ denotes a source term
which creates the $KN$ system with spin-parity $J^{\pi}$ on the lattice.
The four-point correlator in Eq. (\ref{4-point})
is dominated by the lowest energy state with total energy $E_0$
at large time separation ($t \gg t_0 $):
\begin{eqnarray}
{\cal G}_{\alpha}(\vec r, t-t_0; J^{\pi}) 
&=&
\sum_{\vec y}{\cal G}_{\alpha}(\vec x, \vec y, t-t_0; J^{\pi}) \nonumber \\
&\rightarrow&
A_0 \phi_{\alpha}(\vec r; J^{\pi}) e^{-E_0(t-t_0)},
\label{4-point2}
\end{eqnarray}
with $\vec r = \vec x - \vec y$.
Thus, the $KN$ BS wave function is defined by the spatial correlation of
the four-point correlator.
In Eq. (\ref{4-point2}), we assume the Dirichlet boundary condition
in temporal direction, so that the temporal correlation has 
an expnential form, $e^{-E_0(t-t_0)}$.

The BS wave function in s-wave state is obtained under the projection onto 
zero angular momentum ($P^{(l=0)}$),
\begin{equation}
\phi(\vec r; 1/2^-) =
\frac{1}{24}\sum_{g \in O} P^{(l=0)}_{\alpha} 
\phi_{\alpha}(g^{-1} \vec r; 1/2^-),
\label{BS-wave}
\end{equation}
where $g \in O$ represent 24 elements of the cubic rotational group,
and the summation is taken for all these elements.
Using Eq. (\ref{pot}) and Eq. ({\ref{BS-wave}}),
we will find the $KN$ potential and wave function from lattice QCD.

\section{Numerical setup}
In order to calculate the $KN$ potentials in isospin $I=0$ and $I=1$ channels
in 2+1 flavor full QCD,
we have utilized  gauge configurations of JLDG(Japan Lattice Data Grid)/
ILDG(International Lattice Data Grid)
generated by  CP-PACS and JLQCD Collaborations on a $16^3 \times 32$ 
lattice~\cite{Ishikawa:06,JLQCD:07}. 
The renormalization group improved Iwasaki gauge action
and non-perturbatively $O(a)$ improved Wilson quark action are used 
at $\beta=1.83$ which corresponds to the lattice spacing $a=0.1209$ fm
 with the $\rho$ meson mass in the chiral limit.
The physical size of the lattice is about (2.0 fm)$^3$ and the 
the hopping parameters are taken to be 
$\kappa_u=\kappa_d=0.1378$ and $\kappa_s=0.1371$.

In the present simulation, we adopt the spatial wall source located at $t_0$ with
the Dirichlet boundary condition at time slice $t=t_0+16$ in the temporal direction and
the periodic boundary condition in each spatial direction.
The Coulomb gauge fixing is employed at $t=t_0$.
The number of gauge configurations used in the simulation is 700.

\section{Results}

The masses of hadrons are obtained by fitting corresponding two-point correlators.
The obtained masses of the pion, kaon and nucleon are $m_{\pi}=870.7(1.9)$ MeV,
$m_K=911.5(1.9)$ MeV and $m_N=1795.5(6.9)$ MeV, respectively.
Thus the $KN$ threshold energy is 2707 MeV.
\begin{figure}[!htb]
\centering
\includegraphics[width=1.0\columnwidth]{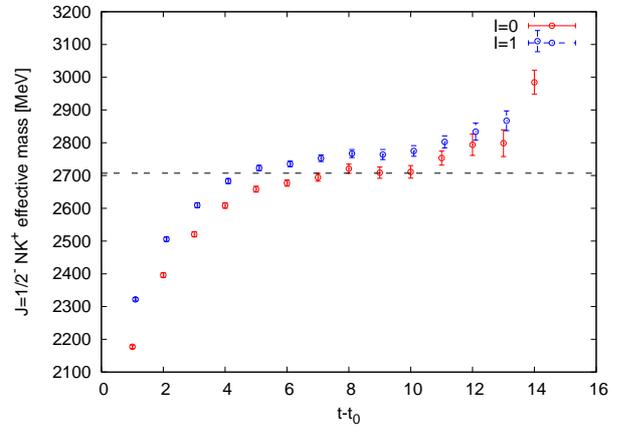}
\caption{Effective mass plot of the $KN$ states in $I=0$(red) and $I=1$(blue) channels.
The dashed line denotes the threshold energy of the $KN$ measured in this simulation.
}
\label{fig:1}       
\end{figure}

Fig.~\ref{fig:1} shows the effective masses of the $KN$ states
in the $I=0$ and $I=1$ channels together with the $KN$ threshold energy.
Effective masses are obtained from the temporal correlation in Eq. (\ref{4-point2}).
We observe plateaus at large time separation, $t-t_0 \ge 7$.
The best fit in the plateau gives $M_{I=0}=2708(11)$ MeV
and $M_{I=1}=2761(10)$ MeV, so that
the lowest $KN$ state is the $I=0$ channel.

\begin{figure}[!htb]
\centering
\includegraphics[width=1.0\columnwidth]{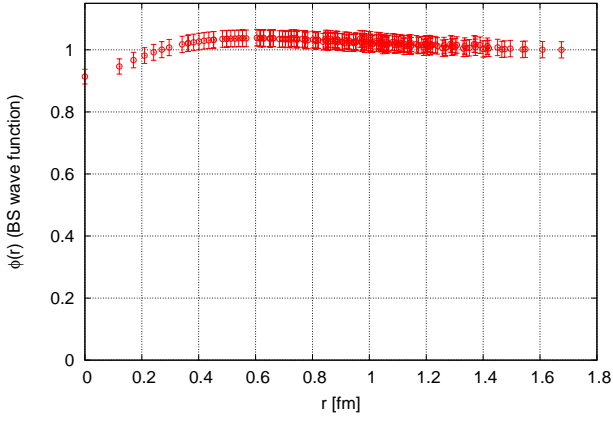}
\caption{The BS wave function of the $KN$ scattering in the $I=0$ channel.
}
\label{fig:2}       
\end{figure}
\begin{figure}[!htb]
\centering
\includegraphics[width=1.0\columnwidth]{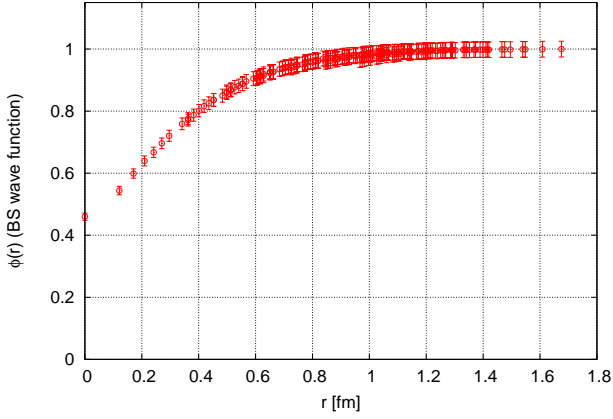}
\caption{The BS wave function of the $KN$ scattering in the $I=1$ channel.
}
\label{fig:3}       
\end{figure}
Figs.~\ref{fig:2} and \ref{fig:3} show BS wave functions of the $KN$ scatterings 
in the $I=0$ and $I=1$ channels, respectively. They are obtained from 
the lattice QCD simulations at large time separation, $t-t_0=8$.
The large $r$ behavior of the BS wave functions in both channels do not show a sign of 
bound state, though more detailed analysis is needed
with large volumes for a definite conclusion.
The small $r$ behavior of the BS wave functions 
suggests some repulsive interaction at short distance ($r<0.3$ fm).
Also, the repulsion 
in the $I=1$ channel seems to be stronger than that in the $I=0$ channel.

\begin{figure}[!htb]
\centering
\includegraphics[width=1.0\columnwidth]{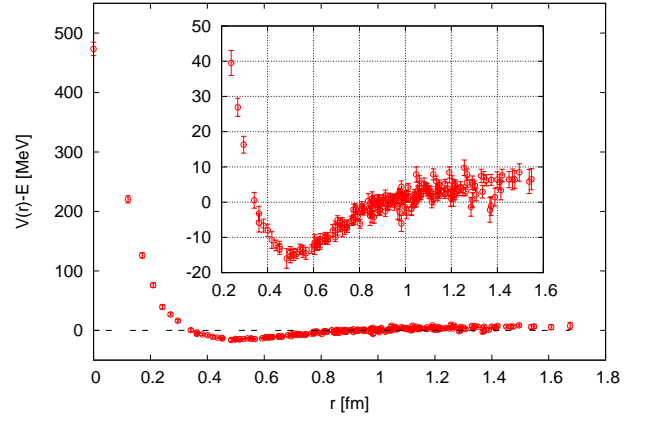}
\caption{The potential of the $KN$ state in the $I=0$ channel without the energy shift
$E$ in Eq. (\ref{pot}).
}
\label{fig:4}       
\end{figure}
\begin{figure}[!htb]
\centering
\includegraphics[width=1.0\columnwidth]{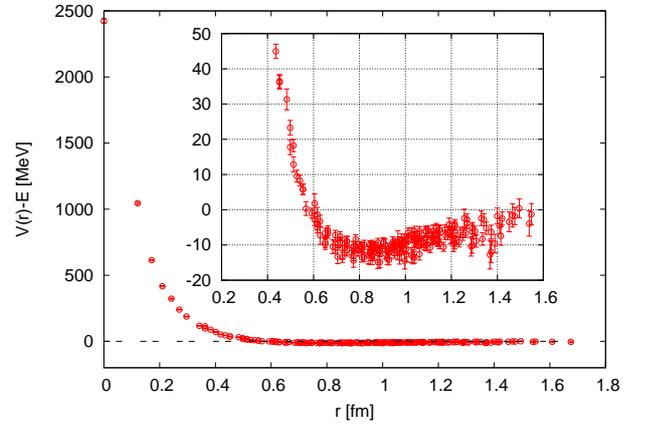}
\caption{The potential of the $KN$ state in the $I=1$ channel without the energy shift 
$E$ in Eq. (\ref{pot}).
}
\label{fig:5}       
\end{figure}

The potential $V(r)$ without the constant energy shift $E$ in Eq. (\ref{pot})
for the $I=0$ ($I=1$) $KN$ state
is shown in Fig.~\ref{fig:4} (Fig. \ref{fig:5}).
These potentials are also calculated by using the data at $t-t_0=8$.
As expected from the BS wave functions in Figs.~\ref{fig:2} and \ref{fig:3},
we observe the repulsive interactions at short distance in both channels.
Also, we observe the attractive well in the 
mid range ($0.4<r<0.8$ fm) in the $I=0$ channel.
In the constituent quark model of hadrons~\cite{Barnes:94},
similar short range repulsion in $KN$ system has been predicted, while
the attraction has not been found.

\section{Discussion}
Results shown in the previous section 
indicate that there are no bound states
in the $I(J^{\pi})=0(1/2^-)$ and $1(1/2^-)$ states 
for the pion mass $m_{\pi} \sim 870$ MeV.

\begin{figure}[!htb]
\centering
\includegraphics[width=1.0\columnwidth]{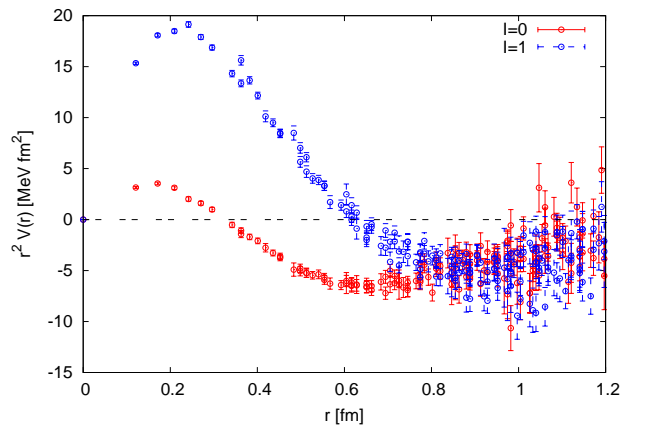}
\caption{The potentials multiplied by a volume factor ($r^2V(r)$).
}
\label{fig:6}       
\end{figure}
In order to compare the strength of the repulsion between
the different isospin states,
we plot the potentials multiplied by a volume factor ($r^2V(r)$) in Fig.~\ref{fig:6}
with the energy shifts $E=-5.0$ MeV for the $I=0$ channel and $E=5.0$ MeV for the
$I=1$ channel.
These energy shifts are estimated from the asymptotic behavior of
the potentials at large $r$,
though analyses with large volumes are needed
to determine these energy shifts more precisely.
As seen in Fig.~\ref{fig:6}, the repulsion at short distance in the $I=1$ channel 
is much stronger than that in the $I=0$ channel.
This is anticipated from the Tomozawa-Weinberg (TW) term in the effective chiral Lagrangian
of mesons and nucleon: The $I=0$ interaction vanishes, 
while the $I=1$ interaction is repulsive from the contact TW interaction, 
which should be compared approximately
with the integral of our $r^2V(r)$ at short distance.

\begin{figure}[!htb]
\centering
\includegraphics[width=1.0\columnwidth]{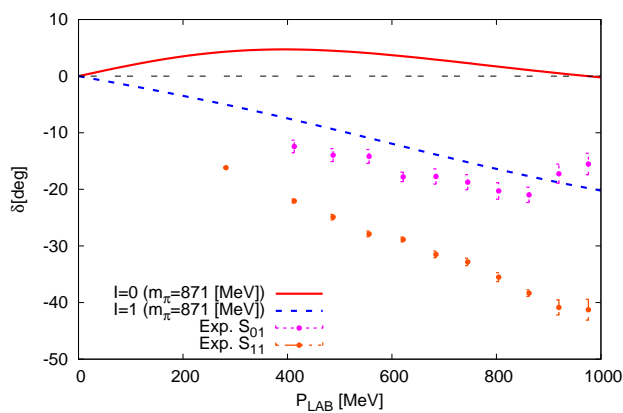}
\caption{The phase shifts of the $KN$ scatterings as the function of 
the momentum ($P_{\rm LAB}$ MeV) in the laboratory frame.
The red solid (blue dashed) curve shows the phase shift of the $I=0$ ($I=1$)
$KN$ scattering. Data are taken from Ref.~\cite{Hashimoto:84}.
}
\label{fig:7}       
\end{figure}

By using the potentials which fit the lattice data in Figs.~\ref{fig:4} and \ref{fig:5},
we can  calculate observables such as the scattering phase shifts.
Fig.~\ref{fig:7} shows such  phase shifts of the $KN$ scattering
 together with the experimental data  
as a function of the laboratory momentum of the kaon.
Theoretical phase shifts are calculated from 
the Schr\"{o}dinger equation with the potentials, $V(r)$, where the energy shift
$E=-5.0$ ($5.0$) MeV in the $I=0$ ($I=1$) channel is taken into account.
Although the hadron masses are  heavy in the present simulation,
qualitative behabior of the phase shifts and also the relative 
magnitude between the $I=0$ and $I=1$ channels are consistent
with the experimental data. 
Simulations along this line with ligher quark masses would eventually
lead to a definite conclusion
on the existence of $\Theta^+$ in $I(J^{\pi})=0(1/2^-)$ and  $1(1/2^-)$
channels.

\section{Summary}
We have performed the 2+1 flavor full lattice QCD simulation to
investigate the $KN$ interaction 
and  a possible signature of the exotic resonance $\Theta^+$.
The s-wave $I=0$ and $I=1$ potentials are extracted from the BS wave functions
for relatively heavy quark mass corresponding to $m_{\pi}=871$ MeV.
Potentials in both channels reveal short range repulsions:
Strength of the repulsion is stronger in the  $I=1$ potential, which is 
consistent with the prediction of the Tomozawa-Weinberg term.
The $I=0$ potential is found to have attractive well at mid range. 
From these potentials, the $KN$ scattering phase shifts are calculated
and compared with the experimental data. Although the quark mass is 
heavy in the present simulation, the results indicate that our method
is promising for future 
quantitative studies of the $KN$ interactions and the exotic resonances
in lighter quark mass region.

\begin{acknowledgement}
We thank Columbia Physics System~\cite{cps} for their lattice QCD simulation code,
of which modified version is used in this work.
The author (Y.I.) thanks Prof. M. Oka for useful discussion.
This work is supported by the Large Scale Simulation Program
No.09-23(FY2009) of High Energy Accelerator Research Organization (KEK),
 the  Grant-in-Aid  of  MEXT (No.20340047)
and  the Grant-in-Aid  for  Scientific Research  on  Innovative  Areas
(No.~2004: 20105001,20105003).

\end{acknowledgement}

\end{document}